# The mutual pulling force of human muscle fibers can treat mild cancer and rhinitis

Hongfa Zi[1], Ding Hua, Zhen Liu[1]

Muscles can store a large amount of genetic information, and in order to transform humans into computers, we need to start by increasing muscle tension. When people with cancer go on happy trips, some cancers often heal without treatment; Rhinitis can cause blockage of the nostrils, but after running, the nostrils naturally ventilate. Both are related to exercise, and the mystery behind them can treat both conditions. Cancer belongs to systemic diseases, and the eradication method for systemic diseases should start from the entire body system, treat the symptoms and prevent recurrence. This article uses special exercise methods and detailed methods to treat diseases, and finds that treating diseases from the perspective of the human system is indeed effective. This article adopts a comparative experimental method to compare the changes in the body before and after. Through this article, it is concluded that exercise and certain methods can cure mild rhinitis and promote rapid ventilation; Explaining from the perspective of muscle pulling force that older individuals are more prone to developing cellular variant cancer; Enhancing muscle tension in the human body can promote the cure of some cancers.

**Keywords:** Human body computer; cellular variation; Muscle pulling force; Mild illness self-healing; Cure of rhinitis

Corresponding author: 1697358179@qq.com

# Introduction

Many cancer patients give up treatment and travel, but the cancer actually disappears. This healing method not only occurs in the field of cancer, but also in the areas of mild angina and brief nasal ventilation, especially in the treatment of rhinitis, which has become a common sense. Gene mutation is the main cause of cancer in many people. The disease is manifested as leukemia and lymphoma (Knudson, 1993), but the highest incidence rate of cancer is not in the muscle area. Obese individuals are more likely to develop cancer, accounting for approximately 20% of the total number of cancers (Wolin et al., 2010). This indicates a certain relationship between muscles and cancer, as well as rhinitis, which is also related to muscles and exercise. Through statistics, it can be observed that both high-intensity and moderate intensity exercise can alleviate rhinitis (Tongtako et al., 2018). Sports can enhance muscle mass and muscle strength. Inflammation such as rhinitis can lead to cellular damage and cancer, which is widely recognized in the medical community. The treatment of symptoms can not only rely on medication, but also on one's own immune system to cure minor illnesses. The human immune system can recognize and attack the source of diseases (Zipfel, 2009); However, the immune system is powerless against cancer cells produced by the human body itself, as cancer cells are a part of the body and cannot be easily recognized and can tolerate the immune system (Pardoll, 2015). But we can increase the muscle content in the body, restore the order of some mutated cells, and more order in the body will lead to fewer cell mutations and cancer. Exercise can increase muscle mass, but excessive exercise can actually lead to muscle damage and inflammation (Appell et al., 1992); Moderate exercise similar to running is more beneficial for overall health and is more economical (Kyröläinen, 2001). So special running methods have better effects on muscle exercise and are easier to popularize. The author will conduct experiments with their own body and explain specific operational and verifiable methods.

Why are elderly people more prone to cancer? What is the relationship between muscle fibers and the treatment of rhinitis and mild cancer in humans? How can we reduce the possibility of cancer? To address the above issues, this article adopts a control experiment method to compare the changes in the body before and after. At the same time, we need specific and bold biological theories to guide the specific experimental process and record it.

This article has three contributions: using exercise and certain methods can cure mild rhinitis and promote rapid ventilation; Explain the reason why older people are more prone to cancer caused by cellular mutations from the perspective of muscle pulling force; Enhancing muscle tension in the human body can promote the cure of some cancers and make mild patients less likely to relapse.

Rhinitis is a very annoying thing, and children with rhinitis do not have the opportunity to get into the top few in learning because it can reduce attention and intelligence in learning, which have been experimentally studied by scientists. I am also a patient with rhinitis, and I am also deeply tormented. But later on, I developed some treatment methods and basically got rid of the torment of rhinitis.

There are still the remaining parts of this article. Observe the research results of previous scholars through a literature review. Then describe the specific experimental method process (without medication), obtain specific experimental results, and analyze the experimental results. Finally, there is the discussion section, which compares it with the research of other scholars.

# Literature Review

After exercise, muscles will appear in the affected area, and the mutual tension between muscles will promote similar cells to be roughly the same. Even if there are different parts of cells, it is difficult to mutate because the interaction between muscles can suppress this variation. The cells that form muscles must be coordinated in order to become tight muscles, and each muscle cell must remain the same to reduce rejection and achieve maximum pulling force.

### Exercise can increase muscle mass and muscle strength

The most common cancer sites in Europe are women's breast cancer, colorectal cancer, lung cancer, and prostate cancer (Ferley and Colombet, 2018). These cancer sites are those with less exercise and less muscle. Many scholars have found that muscle cells seem to inhibit the pathological changes of human cancer cells, reducing the occurrence of many chronic diseases (Pedersen and Febbraio, 2012); The number of muscle cells in the human body is determined at birth, but exercise can increase the diameter and length of muscle fibers (Pearson, 1990). Animals like fish have a higher muscle content than other mammals; Animals that can swim have higher muscle mass than other animals (佐藤健司 et al., 1986). This indicates that streamlined animals can have more muscles, and streamlined muscles in humans are equally important for disease prevention.

### Exercising muscles as a whole to solve overall problems

Many people's cancers become mild or cured after being cured, but the problem of recurrence is still difficult to solve (Esmatabadi et al., 2016). For example, inflammation can lead to more cancer (Coussens and Werb, 2002), so moderate exercise should be chosen instead of intense exercise that can harm the body. The most direct impact of inflammation on the body is rhinitis, which can cause swelling of nasal polyps and then block the nostrils (Ayars, 2000). Inflammation means that the muscle tension in the nose area is gone. High heels have a negative impact on the muscle streamline and function of the legs, so flat shoes should be chosen during exercise (Pannell, 2012). Exercise and rest need to be balanced to form a good cycle. Napping also has great benefits for physical health, as it can increase white blood cell count and enhance human resistance (Faraut et al., 2011)

## Bold Theory

### Older people are more prone to cancer

Elderly people have greater differences in their cells. Human beings constantly divide from fertilized eggs, however, if one cell divides into two cells, there will definitely be differences between these two cells in the end. Split again, the difference is even greater. If differences are placed in all cells of the human body and the cells divide dozens of times, their differences are very significant. The final dividing cell is different from the same type of cell. Cells will lose their ability to divide in the end, and as they age, they lose their ability to coordinate with each other. The loss of order in cells can lead to the loss of some functions. The differences between cells increase, and each cell undergoes extremely weak rejection reactions. The less tension between the muscle fibers of cells, the more relaxed the human skin becomes. That's why people who stay up late and don't exercise are more likely to get cancer. Long term inactivity leads to a decrease in the pulling force of the muscles throughout the body, and prolonged staying up late results in some cells undergoing mutations. This is also the reason why overweight people are more prone to getting sick, because they have more fat and weaker muscle pulling force.

### The impact of body distortion on body cells

Elderly people have greater differences in their cells. Human beings constantly divide from fertilized eggs, however, if one cell divides into two cells, there will definitely be differences between these two cells in the end.

Split again, the difference is even greater. If differences are placed in all cells of the human body and the cells divide dozens of times, their differences are very significant. The final dividing cell is different from the same type of cell. Cells will lose their ability to divide in the end, and as they age, they lose their ability to coordinate with each other. The loss of order in cells can lead to the loss of some functions. The differences between cells increase, and each cell undergoes extremely weak rejection reactions. The less tension between the muscle fibers of cells, the more relaxed the human skin becomes. That's why people who stay up late and don't exercise are more likely to get cancer. Long term inactivity leads to a decrease in the pulling force of the muscles throughout the body, and prolonged staying up late results in some cells undergoing mutations. This is also the reason why overweight people are more prone to getting sick, because they have more fat and weaker muscle pulling force.

## Results

**Taking a nap after running: enhancing muscle fibers and reducing cellular differences**

Due to limited experimental conditions and for practical application, a control experiment method was adopted. In order to obtain more scientific and effective data, pre exercise data (January 9, 2023) was selected as the control group, and the following 5 groups of data were selected as the experimental group (January 10, 2023- January 21, 2023). There are a total of 6 data points, and the observation data was influenced by the duration of exercise.

The number of pimples on the face can measure whether inflammation and cellular variation in the face are enhanced (to a certain extent); 12: 30 and 7:00 body energy are used to measure the vitality brought by midday sleep; The body fatigue at 21:00 is used to measure the degree of physical injury; Mild angina is used to measure the degree of healing of the heart and can reflect the muscle tension in the heart area; Timely ventilation of the nostrils behind the vertical body can reflect the effect of muscle tension on nasal polyps; The sense of suffocation in the nose can reflect the impact of muscle tension on inflammation such as rhinitis.

Special physical exercise. Wear flat shoes for running to reduce leg distortion, especially during exercise. The running time is scheduled at 12:30. Why choose to run at noon instead of in the morning and evening. Because 12 o'clock is the middle of the day at work, cell fatigue caused by labor can also be relieved by sleep (patients do not need to work). The length of running doesn't need to be too long, just run 1000 meters a day. After running, you can sleep for 50-60 minutes, which can reduce the fatigue caused by running. The combination of exercise and sleep means that cells can rest after exercise.

**Get rid of the troubles of rhinitis**

Dredge the nostrils within 5 minutes, and ventilate for a long time. After a few days of exercise, if you have difficulty breathing through your left nostril, you can hunch over and lie in bed. The right shoulder is on the ground, and the left shoulder is not on the ground. Keep the back and shoulders perpendicular, and the head and chest equally perpendicular. And the chin is slightly upward, with the head pressing down on the arm. You will feel a drooping sensation of nasal polyps, and your nose can slowly breathe. This method has a certain purpose: to promote nasal polyps to droop between two nostrils, no longer blocking the nostrils. This method requires a certain level of physical fitness and muscle tone. It does not require surgery to remove nasal polyps (surgery cannot cure them). But we have another question, how to maintain nasal ventilation in a general environment? The above methods are indeed effective, but sitting on a chair for a long time still results in stuffy nostrils. But children need methods to dredge their nostrils while studying. I have found that sitting in a closed and warm environment with a straight waist while reading (books should not be placed flat on the table, but should be tilted at a 30 degree angle) can promote ventilation through the nostrils. Because completely lowering your head to read can cause nasal polyps to fall and block the nostrils (a 30 degree angle is used to keep the face and tabletop slightly perpendicular). Exercise + lying on your side + a 30 degree angle can help dredge your nose, and it can be done continuously throughout the day (brushing your teeth and washing your face with cold water is not recommended). This will not affect the children's learning.

|  | Control group (before exercise) | Experimental group (gradually changing) | | | | |
| --- | --- | --- | --- | --- | --- | --- |
|  | 2023/1/9 | 2023/1/10 | 2023/1/13 | 2023/1/16 | 2023/1/19 | 2023/1/21 |
| The number of pimples on the face (starting at 9:00; the lower the better) | 9 | 9 | 8 | 6 | 5 | 5 |
| Body energy before 12: 30 o'clock (out of 10 points; higher is better) | 6 | 6.5 | 7 | 8 | 8.5 | 8.8 |
| Body energy before 19:00 o'clock (out of 10 points; higher is better) | 5 | 6.5 | 7 | 7.5 | 8.5 | 8.7 |
| Body Fatigue at 21:30 (out of 10 points; lower is better) | 9 | 6 | 5 | 5 | 5 | 5 |
| Mild angina sensation (out of 10 points; lower is better) | 7 | 7 | 6 | 5 | 3 | vanish |
| Is the nostrils properly ventilated after tilting the body (yes or no) | No | No | Yes | Yes | Yes | Yes |
| The suffocation sensation in the nostrils for a day (out of 10 points; the lower, the better) | 9 | 9 | 5 | 5 | 3 | 3 |

Figure 1 shows the comparison before and after exercise

Observing Figure 1, it can be seen that the experimental subjects have fewer pimples and stronger energy. The former means a decrease in inflammation of the face after exercise and an enhancement of facial muscles; The latter means that the experimental subjects have more abundant vitality. This indicates that one's fatigue is significantly reduced, the body suffers less damage at work, and potential inflammation is reduced.

The original mild angina no longer occurs, which means the heart is healthier than before. The nostrils have changed from being unable to ventilate in a timely manner before exercise to being able to ventilate in a timely manner. It represents a gradual increase in muscle tension between nasal polyps and the nose, making the nose healthier. The suffocation sensation in the nostrils is decreasing. It represents a gradual decrease in the recurrence and blockage of rhinitis within a day.

Through experiments, it has been shown that we can increase muscle tone and treat acne on the face through exercise, but elderly people are less likely to exercise too much. It's best not to sweat, as excessive exercise burns life and has the opposite effect. Just take a walk every day, and you should wear shoes with thinner soles when walking.

**Discussion**

The treatment of mild cancer and rhinitis can greatly alleviate people's pain and solve the problem of recurrence of some cancers. Scholars are paying more attention to using surgical knives to achieve cancer cure. The best speed of this method is to solve it quickly, but it has a smaller impact on cancer spread (Mahvi et al., 2018). My experiment focuses more on preventing and enhancing human muscle tension, requiring long-term exercise, and the results are slower. Chemotherapy and taking cancer drugs can target cells at the molecular level, but it can lead to the body becoming increasingly weak (Gilliam and Clair, 2011). My experiment focuses more on methods and tricks, as improving muscle pulling force can enhance vitality. The surgical knife, chemotherapy, and muscle strengthening are not contradictory and can be used in combination. Because as the body of cancer patients weakens with surgery and chemotherapy, enhancing their vitality can help them overcome difficulties.

**Method**

The method used in this article is the control experiment method, which is used to observe whether special exercise is effective for diseases. The control experiment method is mainly used to observe the impact of a certain variable on the results. This article is divided into an experimental group and a control group. In order to facilitate the observation process, the experimental group was divided into 5 groups, and the observation data was influenced by the duration of exercise.

**Data source**

The data comes from the experimental process and has been explained in the text. Relevant personnel can conduct the experiment again. Especially since the experimental process is very simple, it is easier to repeat the experiment. Especially since the experimental process is very simple, it is easier to repeat the experiment.